\documentclass[aps,preprint]{revtex4}%
\usepackage{amsfonts}
\usepackage{amsmath}
\usepackage{amssymb}
\usepackage{graphicx}%
\begin{document}
\preprint{ }
\title{Coordinate Independence and a Physical Metric in Compact Form}
\author{Yukio Tomozawa}
\email{tomozawa@umich.edu}
\affiliation{Michigan Center for Theoretical Physics and Randall Laboratory of Physics,
University of Michigan, Ann Arbor, MI. 48109-1120, USA}
\author{}
\date{\today}

\begin{abstract}
A physical metric is constructed as one that gives a coordinate independent
result for the time delay in infinite order in the perturbation expansion in
the gravitational constant. A compact form for the metric is obtained. One
result is that the metric functions are positive definite. Another is an exact
expression for the gravitational red shift. The metric can be used to
calculate general relativity predictions in higher order for any process. A
relationship between the spacetimes of the physical metric and the
Schwarzschild metric is discussed.

\end{abstract}

\maketitle

\section{\label{sec:level1}Introduction}

It is well known that the coordinates of the Schwartzschild metric do not
correspond to observable physical coordinates. The author has introduced a
method to calculate a coordinate independent result for the time delay for
light propagation in a gravitational field. It has been shown also that there
exists a unique metric that gives the same coordinate independent result and
that the obtained result agrees with the experimental data of Shapiro et. al.
with very good accuracy. For this reason, the author calls this metric a
physical metric. It could also be called a coordinate independent metric or an
invariant metric. In other words, predictions in the physical metric using the
geodesic equation can be compared with observable data without further
coordinate transformation or other adjustments. The consideration of time
delay for light propagation around a mass point is extended to infinite order
in the gravitational constant for the determination of the physical metric and
the coefficients of the perturbation expansion are determined successively.
Summation of the infinite series yields a transcendental equation that leads
to a compact form for the metric functions and can be utilized for expanding
the metric functions at the origin as well as at infinity.. A method for
getting expressions for general relativity predictions in higher order is
presented. An implication of this newly constructed metric is discussed.

\section{Asymptotic form for the physical metric}

The physical metric is expressed as
\begin{equation}
ds^{2}=e^{\nu(r)}dt^{2}-e^{\lambda(r)}dr^{2}-e^{\mu(r)}r^{2}(d\theta^{2}%
+\sin^{2}\theta d\phi^{2}),
\end{equation}
for a spherically symmetric and static mass point $M$. From the fact that the
transformation, $r^{\prime}=re^{\mu(r)/2}$, leads to the Schwartzschild
metric, one can deduce the expression for the metric,%
\begin{equation}
e^{\nu(r)}=1-(r_{s}/r)e^{-\mu(r)/2}, \label{metric1}%
\end{equation}%
\begin{equation}
e^{\lambda(r)}=(\frac{d}{dr}(re^{\mu(r)/2}))^{2}/(1-(r_{s}/r)e^{-\mu(r)/2}),
\label{metric2}%
\end{equation}
where $r_{s}=2GM/c^{2}$ is the Schwartzschild radius. An asymptotic expansion
for the metric functions can be obtained from Eq. (\ref{metric1}) and Eq.
(\ref{metric2}), yielding%
\begin{equation}
e^{\nu(r)}=\sum_{n=0}^{\infty}a_{n}(r_{s}/r)^{n},\text{ }e^{\lambda(r)}%
=\sum_{n=0}^{\infty}b_{n}(r_{s}/r)^{n},\text{ }and\text{\ \ }e^{\mu(r)}%
=\sum_{n=0}^{\infty}c_{n}(r_{s}/r)^{n}, \label{eqasympt}%
\end{equation}
where%
\begin{align}
a_{0}  &  =b_{0}=c_{0}=1,\label{eq3}\\
-a_{1}  &  =b_{1}=1\text{ \ \ }and\label{eq4}\\
a_{2}  &  =c_{1}/2,\text{ \ }b_{2}=1-c_{1}/2+c_{1}^{2}/4-c_{2},\text{ \ }etc.
\label{eq5}%
\end{align}
It is obvious that $a_{n+1\text{ }}$and $b_{n}$ can be expressed as functions
of $c_{n}$, $c_{n-1}$ $\ldots$, $c_{1}$.

\section{Geodesic equations and time delay}

The geodesic equations can be obtained from variations of the line integral
over an invariant parameter $\tau$,$\ \int(\frac{ds}{d\tau})^{2}d\tau$, and
their integrals are given by%

\begin{equation}
\frac{dt}{d\tau}=e^{-\nu(r)}, \label{eq1}%
\end{equation}

\begin{equation}
\frac{d\phi}{d\tau}=J_{\phi}e^{-\mu(r)}/(r\sin\theta)^{2},
\end{equation}

\begin{equation}
(\frac{d\theta}{d\tau})^{2}=(J_{\theta}^{\text{ }2}-J_{\phi}^{\text{ }2}%
/\sin^{2}\theta)e^{-2\mu(r)}/r^{4}.
\end{equation}
Restricting the plane of motion to $\frac{d\theta}{d\tau}=0,$ $\theta=\pi/2,$
the radial part of the geodesic integral is given by%

\begin{equation}
(\frac{dr}{d\tau})^{2}=e^{-\lambda(r)}(e^{-\nu(r)}-J^{\text{ }2}e^{-\mu
(r)}/r^{2}-E) \label{eq2}%
\end{equation}
where $J_{\phi}$, $J_{\theta}$ and $E$ are constants of integration and%

\begin{equation}
J^{\text{ }2}=J_{\phi}^{\text{ }2}=J_{\theta}^{\text{ }2}.
\end{equation}
for the above restriction on the plane of motion. The constant $E$ is 0 for
light propagation.

\bigskip From Eq. (\ref{eq1}) and Eq. (\ref{eq2}) with Eqs. (\ref{eq3}) and
(\ref{eq4}), it follows that%
\begin{align}
\frac{dt}{dr}  &  =\pm\text{ }e^{-\nu(r)}/\sqrt{e^{-\nu(r)-\lambda
(r)}-J^{\text{ }2}e^{-\mu(r)-\lambda(r)}/r^{2}}\\
&  =\pm\text{ }\frac{rr}{\sqrt{r^{2}-r_{0}^{\text{ }2}}}\text{ }%
(1+\frac{(b_{1}-a_{1})\text{ }r_{s}}{2r}+\frac{(c_{1}-a_{1})\text{ }%
r_{0}\text{ }r_{s}}{2\text{ }r\text{ }(r+r_{0})}+\cdots)
\end{align}
for light propagation, where $r_{0}$ is the impact parameter. Integrating from
$r_{0}$ to $r$, one gets the time delay expression for light propagation,%
\begin{equation}
\bigtriangleup t=\text{ }r_{s}\text{ }(\ln(\frac{r+\sqrt{r^{\text{ }2}%
-r_{0}^{\text{ \ }2}}}{r_{0}})+\frac{(c_{1}+1)}{2}\sqrt{\frac{r-r_{0}}%
{r+r_{0}}})+\cdots. \label{eq6}%
\end{equation}
\ 

The second term in Eq. (\ref{eq6}) depends on the choice of the value of
$c_{1}$ and can be eliminated by a further coordinate transformation,%
\begin{equation}
r=r^{\prime\prime}e^{\mu(r^{\prime\prime})/2}=1+c_{1}^{\prime\prime}%
/2(r_{s}/r^{\prime\prime})+\cdots. \label{eq6.5}%
\end{equation}
Therefore, the coordinate independent prediction for time delay in general
relativity should be%
\begin{equation}
\bigtriangleup t=\text{ }r_{s}\text{ }\ln(\frac{r+\sqrt{r^{\text{ }2}%
-r_{0}^{\text{ \ }2}}}{r_{0}})+\cdots\label{eq7}%
\end{equation}
in first order. This is the result also obtained by the PPN (Post Newtonian
Method)\cite{mtw}, and agrees with the most recent observational data
\cite{shapiro 2} with high accurracy (1 in 1000 accurracy). By comparing Eqs.
(\ref{eq6}) and (\ref{eq7}), we conclude that the same coordinate independent
result can be obtained by the condition,
\begin{equation}
c_{1}=-1. \label{eq8}%
\end{equation}

\bigskip We note that the parameter values%
\begin{equation}
a_{1}=-1,\text{ }and\text{ }b_{1}=1 \label{eq8.5}%
\end{equation}
are coordinate independent and determined from the solution of the Einstein
equation and the physical boundary condition. Thus we conclude that Eq.
(\ref{eq8}), along with Eq. (\ref{eq8.5}), is the condition for the physical metric.

\section{The physical metric in higher order}

In order to determine the coefficients in higher order, $c_{n}$, we consider
time delay in the radial direction. For $J=0$, one gets%
\begin{equation}
\frac{dt}{dr}=\text{ }e^{-\nu(r)}/\sqrt{e^{-\nu(r)-\lambda(r)}}=(\frac{d}%
{dr}(re^{\mu(r)/2}))/(1-r_{s}/re^{\mu(r)/2}).
\end{equation}
Integrating this from $r_{1}$ to $r_{2}$, one gets the time difference%
\begin{equation}
\triangle t(r_{1},r_{2})=\int_{r_{1}}^{r_{2}}(\frac{d}{dr}(re^{\mu
(r)/2}))/(1-r_{s}/re^{\mu(r)/2})dr=[re^{\mu(r)/2}+r_{s}\ln(re^{\mu(r)/2}%
-r_{s})]_{r_{1}}^{r_{2}} \label{eq9}%
\end{equation}
Expanding Eq. (\ref{eq9}) in a power series in $r_{s}/r$ (Eq. (\ref{eqasympt}%
), one obtains%
\begin{align}
\triangle t(r_{1},r_{2})  &  =r_{2}-r_{1}+r_{s}\ln(r_{2}/r_{1})+r_{s}%
^{2}(1/r_{2}-1/r_{1})(c_{2}-c_{1}^{2}/4+c_{1}-2)/2+\nonumber\\
&  r_{s}^{3}(1/r_{2}^{2}-1/r_{1}^{2})(c_{3}+c_{2}-c_{1}c_{2}/2+c_{1}%
^{3}/8-c_{1}^{2}/2+c_{1}-1)/2+\cdots. \label{eq10}%
\end{align}
Applying further coordinate transformations, Eq. (\ref{eq6.5}), to Eq.
(\ref{eq10}), one finds the coordinate independent result to be%
\begin{equation}
\triangle t(r_{1},r_{2})=r_{2}-r_{1}+r_{s}\ln(r_{2}/r_{1}) \label{eq11}%
\end{equation}
and the parameters for the physical metric are determined as%
\begin{align}
c_{2}  &  =c_{1}^{2}/4-c_{1}+2=13/4,\nonumber\\
c_{3}  &  =-c_{2}+c_{1}c_{2}/2-c_{1}^{3}/8+c_{1}^{2}/2-c_{1}+1=-9/4,\text{
\ \ }etc., \label{eq12}%
\end{align}
where Eq. (\ref{eq8}) has been used. Successive expansion yields a
determination of all the parameters, $c_{n}$, for the physical metric.

\section{The physical metric in compact form}

From Eq. (\ref{eq9}) and Eq. (\ref{eq11}), it follows that%
\begin{equation}
re^{\mu(r)/2}+r_{s}\ln(re^{\mu(r)/2}-r_{s})-r-r_{s}\ln r=const, \label{q13}%
\end{equation}
where the constant can be determined from the asymptotic form, Eq.
(\ref{eqasympt}), to be%
\begin{equation}
const=(c_{1}/2)r_{s}=-(1/2)r_{s}.
\end{equation}
Then, Eq. (\ref{eq13}) can be tranformed into%
\begin{equation}
f(x)-1+x/2+x\ln(f(x)-x)=0, \label{eq14}%
\end{equation}
where%
\begin{equation}
x=r_{s}/r
\end{equation}
and
\begin{equation}
f(x)=e^{\mu(r)/2}. \label{q14.5}%
\end{equation}
The solution of Eq. (\ref{eq14}) is given by%
\begin{equation}
f(x)=x(1+LambertW(\frac{e^{-(1.5-1/x)}}{x})), \label{eq15}%
\end{equation}
where the Lambert W function is the solution of the equation\cite{lambert},%
\begin{equation}
W(z)e^{W(z)}=z. \label{eq16}%
\end{equation}

The compact form for the angular metric function in the physical metric, Eq.
(\ref{eq14}) or Eq. (\ref{eq15}) can be used to calculate an asymptotic
expansion or an expansion at the origin. A tedious but doable calculation
yields%
\begin{align}
re^{\mu(r)/2}  &  =rf(r_{s}/r)\\
&  =r(1-\frac{1}{2}(r_{s}/r)+\frac{3}{2}(r_{s}/r)^{2}-\frac{3}{8}(r_{s}%
/r)^{3}+\cdots) \label{eq17}%
\end{align}
for the asymptotic form and%
\begin{align}
re^{\mu(r)/2}  &  =r_{s}(1+\alpha_{0}(r/r_{s}+(1-\alpha_{0})((r/r_{s}%
)^{2}+\frac{1}{2}(1-3\alpha_{0})(r/r_{s})^{3}+\cdots)))\label{eq18}\\
&  =r_{s}(1+.2231301601(r/r_{s})+.1733430918(r/r_{s})^{2}+.02865443810(r/r_{s}%
)^{3}+\cdots) \label{eq19}%
\end{align}
for the expansion at the origin, where%
\begin{equation}
\alpha_{0}=e^{-\frac{3}{2}}=.2231301601. \label{eq20}%
\end{equation}

Using Eqs. (\ref{metric1}), (\ref{metric2}) and (\ref{eq17}-\ref{eq20}), one
can express the metric functions in asymptotic form and expanded at the
origin,%
\begin{align}
e^{\nu(r)}  &  =1-r_{s}/re^{\mu(r)/2}=1-r_{s}/r-\frac{1}{2}(r_{s}/r)^{2}%
+\frac{5}{4}(r_{s}/r)^{3}+\cdots\label{eq20.1}\\
&  =\alpha_{0}(r/r_{s})(1+(1-2\alpha_{0})(r/r_{s})+\frac{1}{2}(1-8\alpha
_{0}+9\alpha_{0}^{2})(r/r_{s})^{2}+\cdots)\\
&  =.2231301601(r/r_{s})+.1235560234(r/r_{s})^{2}-.03759270901(r/r_{s}%
)^{3}+\cdots, \label{eq20.5}%
\end{align}%
\begin{align}
e^{\lambda(r)}  &  =(\frac{d}{dr}(re^{\mu(r)/2}))^{2}/(1-r_{s}/re^{\mu
(r)/2})=1+r_{s}/r-\frac{3}{2}(r_{s}/r)^{2}-\frac{3}{4}(r_{s}/r)^{3}+\cdots\\
&  =\alpha_{0}(r_{s}/r)(1+(3-2\alpha_{0})(r/r_{s})+\frac{1}{2}(7-16\alpha
_{0}+9\alpha_{0}^{2})(r/r_{s})^{2}+\cdots)\\
&  =.2231301601(r_{s}/r)+0.5698163437+0.4326494982(r/r_{s})+\cdots
\end{align}%
\begin{align}
e^{\mu(r)}  &  =(\frac{re^{\mu(r)/2}}{r})^{2}=1-(r_{s}/r)+\frac{13}{4}%
(r_{s}/r)^{2}-\frac{9}{4}(r_{s}/r)^{3}+\cdots\\
&  =(r_{s}/r)^{2}(1+\alpha_{0}(2r/r_{s}+(2-\alpha_{0})((r/r_{s})^{2}%
+(1-\alpha_{0})^{2}(r/r_{s})^{3}+\cdots)))\\
&  =(r_{s}/r)^{2}+.4462603202(r_{s}/r)+.3964732520+.1346650199(r_{s}%
/r)+\cdots.
\end{align}

\section{The nature of the physical metric}

The metric functions for the physical metric and the coordinate transformation
from the Schwarzschild metric, Eq. (\ref{eq17}), are now shown to be positive
definite. In order to show this property, the solution of Eq. (\ref{eq14})
must satisfy the condition,%
\begin{equation}
f(x)\geqq x, \label{eq29}%
\end{equation}
or equivalently%
\begin{equation}
re^{\mu(r)/2}\geqq r_{s}. \label{eq30}%
\end{equation}
The equality in Eq. (\ref{eq30}) is valid only at $r=0,$as is seen in Eq.
(\ref{eq19}). From this condition, it follows that%
\begin{equation}
e^{\nu(r)}=1-(r_{s}/r)e^{-\mu(r)/2}\geqq0, \label{eq31}%
\end{equation}
where equality is valid only at the origin, $r=0$. As a result, the positive
definiteness of $e^{\lambda(r)}$ follows easily,%
\begin{equation}
e^{\lambda(r)}=(\frac{d}{dr}(re^{\mu(r)/2}))^{2}/(1-(r_{s}/r)e^{-\mu(r)/2})>0.
\label{eq32}%
\end{equation}

A further relationship among the metric functions follows from the generation
equation, Eq. (\ref{eq14}). Rewriting it as%
\begin{equation}
re^{\mu(r)/2}-r+r_{s}/2+r_{s}\ln((re^{\mu(r)/2}-r_{s})/r)=0,
\end{equation}
and differentiating it yields the relationship,%
\begin{equation}
(re^{\mu(r)/2})^{%
\acute{}%
}=(1-(r_{s}/r)e^{-\mu(r)/2})(1+r_{s}/r)=e^{\nu(r)}(1+r_{s}/r). \label{eq33}%
\end{equation}
Hence one gets%
\begin{equation}
e^{\lambda(r)}=(1-(r_{s}/r)e^{-\mu(r)/2})(1+r_{s}/r)^{2}=e^{\nu(r)}%
(1+r_{s}/r)^{2}. \label{eq34}%
\end{equation}
This confirms the positive definiteness of $e^{\lambda(r)}$. From%
\begin{equation}
(e^{\nu(r)})^{%
\acute{}%
}=\frac{r_{s}}{(re^{\mu(r)/2})^{2}}(re^{\mu(r)/2})^{%
\acute{}%
}=\frac{r_{s}}{(re^{\mu(r)/2})^{2}}e^{\nu(r)}(1+r_{s}/r) \label{eq35}%
\end{equation}
and Eq. (\ref{eq33}), it follows that $e^{\nu(r)}$ and $re^{\mu(r)/2}$ are
monotonically increasing functions with $0\leqq$ $e^{\nu(r)}\lessdot1$ and
$re^{\mu(r)/2}\gtrdot r_{s}$. From Eq. (\ref{eq20.1}) and Eq. (\ref{eq20.5}),
it is easily seen that there should exist an inflection point for $e^{\nu(r)}%
$. By a numerical computation, it can be shown that $e^{\nu(r)}$ has a single
inflection point at $r=0.65926r_{s}$.

It is now clear that the metric functions in the physical metric are positive
definite. The time metric function, $g_{00}=$ $e^{\nu(r)}$, vanishes at the
origin. In other words, there is no horizon at finite distance in this metric.
One may say that the horizon coincides with the origin. This is not
surprising, since there is no trace of a horizon in the invariants made from
the Riemann curvature tensors, $R_{\alpha\beta\gamma\delta}R^{\alpha
\beta\gamma\delta}$. In the Schwarzschild metric, the latter is $12r_{s}%
^{2}/r^{\prime6}$. Therefore, in the physical metric one obtains%
\begin{equation}
R_{\alpha\beta\gamma\delta}R^{\alpha\beta\gamma\delta}=12r_{s}^{2}%
/(re^{\mu(r)/2})^{6}.
\end{equation}
This quantity neither vanishes nor has a singulaity at a finite distance. The
author will come back to this issue in the last discussion section.

\section{Gravitational red shift}

In the physical metric, the formula for the gravitational red shift is
obtained from the proper time expression,%
\begin{align}
d\tau_{p}  &  =e^{\nu(r)/2}dt\\
&  =\sqrt{1-r_{s}/r-\frac{1}{2}(r_{s}/r)^{2}+\frac{5}{4}(r_{s}/r)^{3}+\cdots
}dt. \label{eq21}%
\end{align}
Since $e^{\nu(r)}$ is positive definite, Eq. (\ref{eq21}) is valid for all
values of r, in contradistinction to the case of the Schwarzschild metric. Of
course, near the origin, the expansion at the origin, Eq. (\ref{eq20.5}),
should be used. It is worthwhile to point out that Eq. (\ref{eq21}) is the
first general relativity prediction in higher order for the gravitational red
shift, which can be utilized for experimental tests in the future. How likely
is it to be used in the near future? The values of the parameter, $r_{s}/r$,
at the surface of a white dwarf and the sun are $10^{-3}$ and $10^{-5}$
respectively. It is, therefore, not wild imagination that second order effects
could be detected in gravitational red shifts in atomic spectra from white
dwarfs or the sun sometime in the near future.

\section{Bending of light in higher order}

The bending of light can be obtained from%
\begin{align}
\pi+\triangle\phi &  =2\int(e^{-\mu(r)+\lambda(r)/2}/r^{2}\sqrt{e^{-\nu
(r)}/J^{2}-e^{-\mu(r)}/r^{2}})dr\\
&  =2\int_{r_{0}}^{\infty}\frac{d}{dr}(re^{\mu(r)/2})dr/(re^{\mu(r)/2}%
)^{2}\sqrt{1/J^{2}-(1-r_{s}/(re^{\mu(r)/2}))/(re^{\mu(r)/2})^{2}}.
\end{align}
Changing the integration variable to%
\begin{equation}
u=1/(re^{\mu(r)/2}),
\end{equation}
one obtains%
\begin{equation}
\pi+\triangle\phi=2\int_{0}^{u_{0}}du/\sqrt{u_{0}^{2}(1-r_{s}u_{0}%
)-u^{2}(1-r_{s}u)},
\end{equation}
where%
\begin{equation}
u_{0}=1/(r_{0}e^{\mu(r_{0})/2}).
\end{equation}
Defining%
\begin{equation}
u=u_{0}s
\end{equation}
and%
\begin{equation}
k=r_{s}u_{0}=r_{s}/(r_{0}e^{\mu(r_{0})/2}),
\end{equation}
one gets%
\begin{equation}
\pi+\triangle\phi=2\int_{0}^{1}ds/\sqrt{(1-s^{2})(1-\kappa(\frac{1}{1+s}+s)}.
\label{eq40}%
\end{equation}
Expanding in a power series in $\kappa$, one gets%
\begin{equation}
\triangle\phi=2\sum_{n=1}^{\infty}\frac{(2n-1)!!}{2^{n}n!}\kappa^{n}\int
_{0}^{1}\frac{1}{\sqrt{1-s^{2}}}(\frac{1}{1+s}+s)^{n}.
\end{equation}

Defining the integrals,
\begin{align}
I_{n}  &  =\int_{0}^{1}\frac{1}{\sqrt{1-s^{2}}}(\frac{1}{1+s}+s)^{n}ds,\\
A_{n}  &  =\int_{0}^{1}\frac{1}{\sqrt{1-s^{2}}}(\frac{1}{1+s})(\frac{1}%
{1+s}+s)^{n}ds
\end{align}
and%
\begin{equation}
B_{n}=\int_{0}^{1}\frac{s}{\sqrt{1-s^{2}}}(\frac{1}{1+s}+s)^{n}ds,
\end{equation}
one gets recursion formulas%
\begin{align}
I_{n}  &  =A_{n-1}+B_{n-1},\\
A_{n}  &  =\frac{1}{2n+1}(1-nI_{n}+3nI_{n-1})
\end{align}
and%
\begin{equation}
B_{n}=\frac{1}{n+1}(1-3nA_{n-1}+3nI_{n-1}).
\end{equation}
Using%
\begin{equation}
A_{0}=B_{0}=1,
\end{equation}
and%
\begin{equation}
I_{0}=\frac{\pi}{2},
\end{equation}
repeated application of the recursion formulas yields%
\begin{equation}
I_{1}=2,\text{ }I_{2}=-\frac{4}{3}+\frac{5}{4}\pi,\text{ }I_{3}=\frac{122}%
{15}-\frac{3}{2}\pi,\text{ }I_{4}=-\frac{104}{7}+\frac{99}{16}\pi,\text{
}\cdots.
\end{equation}
With the additional use of the expression%
\begin{align}
\kappa &  =r_{s}/(r_{0}e^{\mu(r_{0})/2})\\
&  =(r_{s}/r_{0})/(1-\frac{1}{2}(r_{s}/r_{0})+\frac{3}{2}(r_{s}/r_{0}%
)^{2}-\frac{3}{8}(r_{s}/r_{0})^{3}-\frac{3}{4}(r_{s}/r_{0})^{4}+\frac{21}%
{64}(r_{s}/r_{0})^{5}+\frac{291}{320}(r_{s}/r_{0})^{6}+\cdots)\\
&  =(r_{s}/r_{0})(1+\frac{1}{2}(r_{s}/r_{0})-\frac{5}{4}(r_{s}/r_{0}%
)^{2}-(r_{s}/r_{0})^{3}+\frac{37}{16}(r_{s}/r_{0})^{4}+\frac{143}{64}%
(r_{s}/r_{0})^{5}-\frac{379}{80}(r_{s}/r_{0})^{6}+\cdots),
\end{align}
one gets a formula for higher order corrections to the bending of light,%
\begin{equation}
\triangle\phi=2(r_{s}/r_{0})+\frac{15}{16}\pi(r_{s}/r_{0})^{2}+\frac{19}%
{12}(r_{s}/r_{0})^{3}-(\frac{1}{4}+\frac{135}{1024}\pi)(r_{s}/r_{0}%
)^{4}+\cdots.
\end{equation}

The second order correction term is consistent with that obtained in the
references\cite{2ndorder} that use the impact parameter method. It is an
interesting question whether the two methods, one using the physical metric
and the other using the impact parameter, give the same result for higher
order corrections. If they are different, the natural question is which method
gives the correct answer. Of course, it is difficult to answer such a question
based on the experimental data for general relativity tests, since one
requires high precision experiments to do that. It is also an interesting
question whether the impact parameter method yields a result that implies the
absence of a horizon, as was shown for the physical metric method.

\section{Period of light in a circular orbit}

In this section, it is shown that the period of light in a circular orbit is
coordinate independent.\cite{davies} From Eq. (\ref{eq40}), it is clear that
the integral diverges if the value of $\kappa$ approaches%
\begin{equation}
\kappa=\frac{2}{3}.
\end{equation}
In other words, light makes a circular orbit at%
\begin{equation}
\kappa=r_{s}/(re^{\mu(r)/2})=\frac{2}{3}. \label{eq41}%
\end{equation}
Using the fact that light in a circular orbit satisfies the equation%
\begin{equation}
\frac{dt}{d\phi}=\frac{re^{\mu(r)/2}}{e^{\nu(r)/2}}=\frac{re^{\mu(r)/2}}%
{\sqrt{1-\frac{r_{s}}{re^{\mu(r)/2}}}}=\frac{3\sqrt{3}}{2}r_{s},
\label{eq41.1}%
\end{equation}
the period for circular motion is given by%
\begin{equation}
T=\int_{0}^{2\pi}\frac{dt}{d\phi}d\phi=2\pi\frac{3\sqrt{3}}{2}r_{s}.
\label{eq41.5}%
\end{equation}
This result is obviously coordinate independent.

The solution of Eq. (\ref{eq41}) yields the radius of the circular orbit for
the particular metric. For the Schwarzschild metric $(e^{\mu(r)/2}=1)$, one
has%
\begin{equation}
r=\frac{3}{2}r_{s},
\end{equation}
while for the Eddington metric $(e^{\mu(r)/2}=(1+r_{s}/4r)^{2})$, one gets%
\begin{equation}
r=\frac{(2+\sqrt{3})}{4}r_{s}.
\end{equation}
For the physical metric, from Eq. (\ref{eq41}) it follows that%
\begin{equation}
f(x)=\frac{3}{2}x. \label{eq42}%
\end{equation}
Combining Eq. (\ref{eq14}) and Eq. (\ref{eq42}), one gets the equation%
\begin{equation}
e^{2-1/x}=2/x,
\end{equation}
and the solution%
\begin{equation}
r=1.15916r_{s}.
\end{equation}
The meaning of Eqs. (\ref{eq41.1}) and (\ref{eq41.5}) may be understood as%
\begin{equation}
T=\frac{2\pi r}{e^{\nu(r)/2}/e^{\mu(r)/2}}=\frac{2\pi r}{(\text{speed of
light})},
\end{equation}
where%
\begin{equation}
(\text{speed of light})=\frac{rd\phi}{dt}=e^{\nu(r)/2}/e^{\mu(r)/2}%
\end{equation}
is measured by a clock at infinite distance.

\section{Discussion}

What is the meaning of the physical metric other than that it gives coordinate
invariant results and that it gives a prediction consistent with observations,
at least in first order of the gravitational constant? The reason for the
positive definiteness of the metric functions of the physical metric is
understood easily: The transformation from r to the Schwarzschild coordinate,%
\begin{equation}
r^{\prime}=re^{\mu(r)/2}, \label{eq43}%
\end{equation}
or rather the inverse of it is a map of the outside of the Schwarzschild
spacetime onto the whole space of the physical metric. This can be seen also
from Eqs. (\ref{eq18}) or (\ref{eq19}) and%
\begin{equation}
\lim_{r\rightarrow0}(re^{\mu(r)/2})=r_{s}. \label{eq43.1}%
\end{equation}
In other words, Eq. (\ref{eq43}) is a map from the physical metric space to
the outside of the Schwarzschild radius. It is not intended, but the
requirement of coordinate indepence forces us to choose this mapping. One can
present two possible interpretations of this result.

[A]. An observer outside the horizon (the Schwarzschild radius) can see only
the outside, but cannot see inside the horizon, since the speed of light
vanishes at the horizon and nothing can come out from the horizon in the usual
interpretation of black holes. To the extent that the requirement of
coordinate independence leads to physically observable consequences, (which
has been proved only in first order of the gravity,) the nature of the map
between the physical metric space and the Schwarzschild spacetime, Eqs.
(\ref{eq43}) and (\ref{eq43.1}), is comprehensible. In order to see the whole
picture of the spacetime, one has to make a map that opens up the part of the
spacetime inside the horizon.

[B]. The spacetime of the physical metric is a real coordinate system that
describes nature, since that is all one can get by observations. The spacetime
that is opened up by a coordinate transformation is an artifact of the
coordinate transformation. In this view, the horizon and anything that comes
out from it has no reality. That includes Hawking radiation, time tunneling
and the information paradox.

A response from the viewpoint [A] against the viewpoint [B] could be that
sometimes poles in the complex plane of a physical quantity has physical
reality. In a S-matrix theory, poles in the unphysical region for physical
parameter spces correspond to observable particles or bound states. Then, the
viewpoint [B] would say that yes, then one has to discover a horizon as a
physical observable quantity. And a dispute continues.

The only way to resolve these two viewpoints is to find direct or indirect
observational evidences to support its viewpoint. In order to verify the
viewpoint [A], an observation of Hawking radiations is essential. Measurements
of the Hawking radiations from primordial black holes are inportant projects
that \ can prove viewpoint [A]. Or one may create a method to find an evidence
for the existence of horizon from some other phenomena. So far, evidences for
black holes in quasars or active galactic nuclei come from phenomena related
to acretion disks around massive objects. The existence of compact massive
objects does not differentiate the both viewpoint, [A] or [B], since such a
phenomena can be claimed from the both viewpoints.

\begin{acknowledgments}
It is a great pleasure to thank David N. Williams for reading the manuscript
and Paul C. W. Davies, Jean Krisch and David N. Williams for helpful discussions.
\end{acknowledgments}


\begin{thebibliography}{9}
\bibitem {davies}The work in this section is due to a suggestion and a
remarkable intuition of Paul Davies.
\end{thebibliography}
\end{document}